\documentclass[final,5p,times,twocolumn]{elsarticle} 
\usepackage{epsfig}  
\usepackage{amssymb,amsmath}
\usepackage{tabularx}
\usepackage[table,xcdraw]{xcolor}
\usepackage{natbib}
\usepackage[greek,english]{babel}
\usepackage{subcaption}
\usepackage{float}
\usepackage[colorlinks=true,linkcolor=blue,urlcolor=blue]{hyperref}
\usepackage{ulem}

\journal{Physica A: Statistical Mechanics and its Applications}

\begin{document}

\begin{frontmatter} 
\title{
A mean field analysis of the role of indirect transmission in emergent infection events}

\author[tomas]{Tomás Ignacio González\corref{cor1}}
\ead{tomignaciogon@gmail.com}
\cortext[cor1]{Corresponding author}

\author[fabiana]{María Fabiana Laguna}
\ead{lagunaf@cab.cnea.gov.ar}

\author[guillermo]{Guillermo Abramson}
\ead{abramson@cab.cnea.gov.ar}

\address[tomas]{Statistical and Interdisciplinary Physics Division, Centro Atómico Bariloche (CNEA), R8402AGP Bariloche, Argentina.}

\address[fabiana]{Statistical and Interdisciplinary Physics Division, Centro Atómico Bariloche (CNEA) and CONICET. Universidad Nacional de Río Negro. R8402AGP Bariloche, Argentina.}

\address[guillermo]{Statistical and Interdisciplinary Physics Division, Centro Atómico Bariloche (CNEA), CONICET and Instituto Balseiro (Universidad Nacional de Cuyo). R8402AGP Bariloche, Argentina.}

\begin{abstract}
We developed a mathematical model to investigate the role of indirect transmission in the spread of infectious diseases, using the illustrative example of sarcoptic mange as a case study.
This disease can be transmitted through direct contact between an infected host and a susceptible one, or indirectly when potential hosts encounter infectious mites and larvae deposited in the environment, commonly referred to as fomites. Our focus is on exploring the potential of these infectious reservoirs as triggers for emerging infection events and as stable reservoirs of the disease. To achieve this, our mean field compartmental model incorporates the epidemiological dynamics driven by indirect transmission via fomites.
We identify  different types of dynamics that the system can go into, controlled by different levels of direct and indirect transmission. Among these, we find a new regime where the disease can emerge and persist over time solely through fomites, without the necessity for direct transmission. This possibility of the system reveals an evolutionary pathway that could enable the parasite to enhance its fitness beyond host co-evolution. We also define a new threshold based on an effective reproductive number, that enables us to predict the conditions for disease persistence. 
Our model allows us to assess the potential effectiveness of various disease intervention measures by incorporating a feature observed in real systems. We hope this contributes to a better understanding of infectious disease outbreaks.

\end{abstract}

\begin{keyword}
mange, fomite, indirect transmission, mathematical epidemiology
\end{keyword}

\date{\today}
\end{frontmatter}

\section{Introduction}

The representation and analysis of epidemics through mathematical modeling provides quantitative descriptions of the relationships between variables and enables the formulation and testing of hypotheses. 
By identifying the parametric conditions associated with specific behaviors of the system, the mathematical framework can be useful in understanding the different mechanisms at play, as well as making predictions about the global dynamics of an epidemic \citep{grassly2008mathematical, garnett2011mathematical}. Of particular interest are emergent infectious diseases, caused by unknown pathogens or by new variants of previously studied ones \citep{mills2006biodiversity,williams2002emerging}. These diseases currently constitute a severe threat, not only to human populations, but also to biodiversity and ecological networks in various parts of the world. Their effects can range from increases in natural mortality rates to local extinction of native species \cite{daszak2000emerging,lafferty2002good}.

Among the infectious agents we can count not only microbial and viral pathogens \cite{schmaljohn97,abramson2002,laneri2021}, but also eukaryote parasites, like protozoa, helminths and arthropods \cite{laneri2015,laguna2011}. Within this last group, we can mention \textit{Sarcoptes scabiei}, which is considered an important threat for wildlife, with a great impact on a wide range of host species and ecosystems \citep{pence2002sarcoptic,thompson2009parasite,astorga2018international, ferreyra2022sarcoptic, monk2022}. The disease associated with \textit{Sarcoptes scabiei} is known as  sarcoptic mange (or scabies in the case of humans), and its symptoms originate as a hypersensitive allergic response from the host, which can lead to a chronic process, with a  decrease of the host's performance \citep{ferreyra2022sarcoptic, bornstein2002sarcoptes,rojas2004nosoparasitosis,niedringhaus2019review,arlian2017review}. The present work has been inspired by an outbreak of sarcoptic mange in wild camelids populations, that occurred in 2014 in San Juan, Argentina \citep{ferreyra2022sarcoptic}. Nevertheless, we keep our treatment at an abstract level, providing a general framework for the description of similar situations. 

During the International Meeting on Sarcoptic Mange in Wildlife in 2018, six key points of interest were identified to further understand this infectious disease, including its natural dynamics and various forms of transmission within wild populations. This is due to the fact that, despite the empirical evidence emphasizing the role of sarcoptic mange as an emergent threatening disease for wildlife \cite{monk2022}, the causes of the development of these epidemic episodes are still poorly understood \citep{pence2002sarcoptic,astorga2018international}. 

\textit{Sarcoptes scabiei}  constitutes a case of a parasite that can be transmitted through direct contact between a current host and a potential one, or indirectly when a potential host comes into contact with infectious mites and larvae deposited in the environment, commonly called \textit{fomites}. The role of both types of contact may vary for different susceptible species \citep{astorga2018international,niedringhaus2019review,arlian1989biology}. 
The post-invasion dynamics in wild populations have been described as variable, resulting in either local declines or endemic states of the disease \cite{ferreyra2022sarcoptic,monk2022,kerlin2022,kerlin2022b}. Additionally, evidence suggests that these outcomes could stem from both direct, density-dependent dynamics, and indirect, density-independent transmission \citep{martin2019population,carver2023density}.

Taking sarcoptic mange as a reference parasite model, our aim is to analyze the potential of fomites as a triggering factor for emerging events and also as a reservoir of the disease, capable of persisting over time. For this, we have developed a mean field system based on classical epidemiological models \citep{anderson1991discussion, blower2008modelling}, with the addition of a compartment that represents the fomites as an alternative source of infection \cite{yagci2019comparing}. This increase in the analytical complexity of the system is rewarded with a more comprehensive, closer to reality, perspective on the disease dynamics, providing a better understanding for eventual management decisions \citep{chubb2010mathematical,chowell2016mathematical}.

\section{Model description}
To study and describe the mean field dynamic of sarcoptic mange infested population, taking into account both direct and indirect transmission, we developed a system of differential equations based on classical compartmental models
\citep{anderson1991discussion, blower2008modelling}. 
Specifically, our model is an extension of an SEI model, with vital dynamics represented by a logistic term. The rationale for incorporating this term is twofold: firstly, the time scale associated with sarcoptic mange outbreaks is comparable to the lifespan of the host population, resulting in the establishment of endemic states \citep{martin2019population, carver2023density}; secondly, its inclusion alters the estimation and characterization of the system's fixed points, defining threshold values and thus generating new potential scenarios for the development of the disease \citep{abramson2002,kermack1932contributions, gomes2009sir}. The SEI model is represented by the following equations:

\begin{align}
\frac{dS}{dt} &= r(S+E)(1-S-E-I) - \beta_1 SI, \label{eq0S}\\
\frac{dE}{dt} &= \beta_1 SI  - \gamma E, \label{eq0E}\\
\frac{dI}{dt} &= \gamma E - \mu I. 
\label{eq0I}
\end{align}
Each equation reflects the dynamic of one of the epidemiological compartments: $S$ are the healthy and susceptible individuals, $E$ (exposed) are the non-infectious carriers of the disease, and $I$ are the infectious individuals, capable of transmitting the disease. The parameters of this reference model are $r$ (net reproduction rate), $\beta_{1}$ (direct transmission rate of the disease), $\gamma$ (rate at which the non-infectious carriers become infectious), and $\mu$ (death rate of the infected individuals) \cite{keeling2011modeling}. 
In this foundational model, transmission solely occurs through infected individuals, without considering the presence of fomites or any other form of indirect transmission pathways.

\begin{figure}[h]
\centering
\includegraphics[width=\columnwidth]{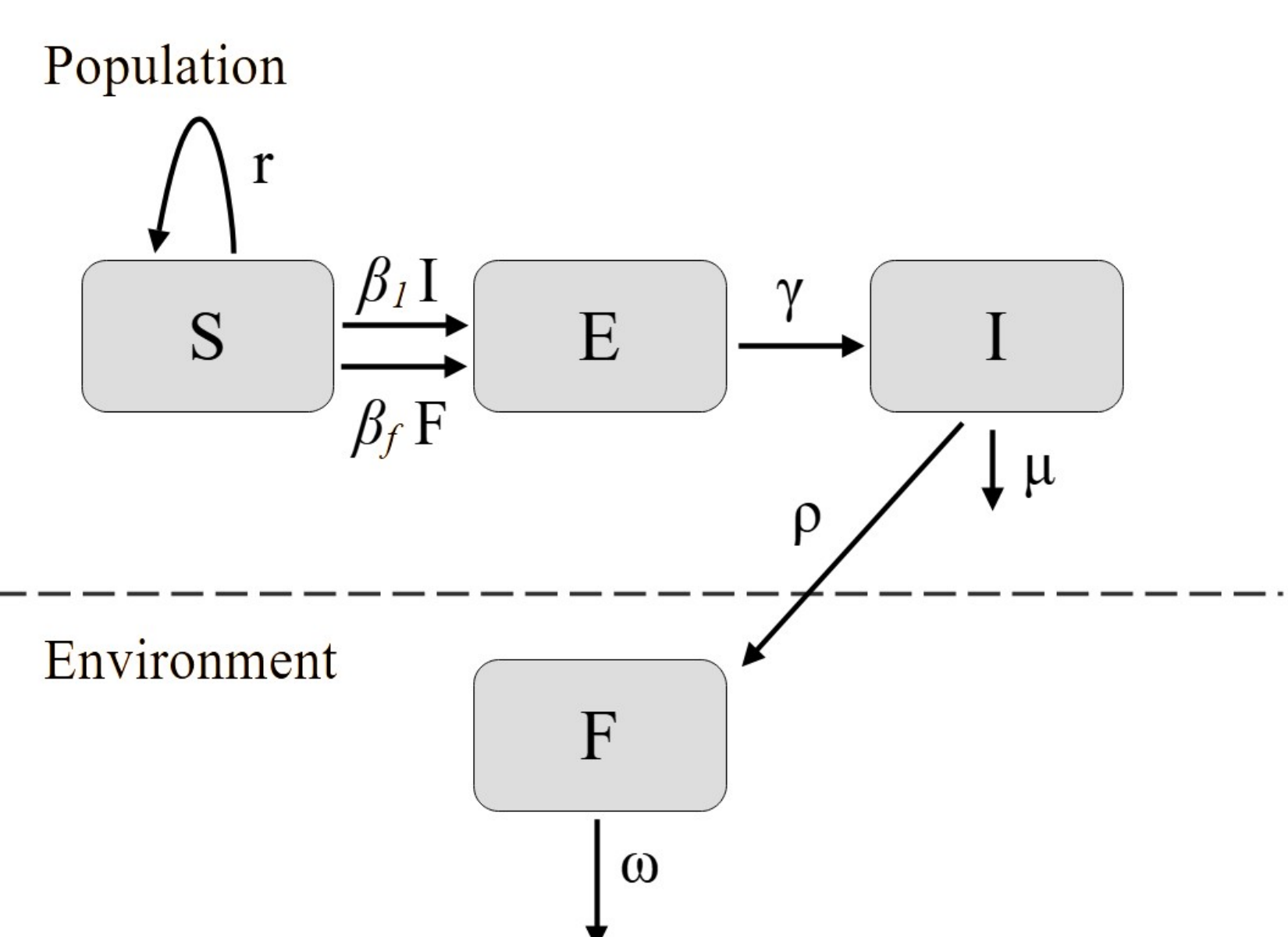}
\caption{Scheme of the SEIF model}
\label{fig:fig0}
\end{figure}

In our proposal, we introduce a fourth component that describes the dynamic of fomites, $F$, together with corresponding new terms in Eqs.~(\ref{eq0S}) and~(\ref{eq0E}), accounting for the indirect transmission: 
\begin{align}
\frac{dS}{dt} &= r(S+E)(1-S-E-I) - \beta_1 SI - \beta_f SF, \label{eq1S}\\
\frac{dE}{dt} &= \beta_1 SI + \beta_f SF - \gamma E, \label{eq1E} \\
\frac{dI}{dt} &= \gamma E - \mu I, \label{eq1I} \\
\frac{dF}{dt} &= \rho I - \omega F. 
\label{eq1F}
\end{align}
The rate at which these reservoirs are introduced into the environment is determined by the parameter $\rho$ and the total number of infected individuals in the system. The rate at which they are removed from the system is given by $\omega$, the inverse of the maximum survival time of eggs and larvae outside the host. Finally, the indirect transmission of the disease over the susceptible is quantified by the parameter $\beta_{f}$, which constitutes an additional non-linear source of exposure in the dynamics (Fig.~\ref{fig:fig0})$^2$.
Similar treatments to include the mechanism of indirect transmission were conducted in studies \citep{yagci2019comparing,cortez2013distinguishing}, although both the initial models and their approach are different from what we have done in this work.

Another distinctive aspect of our model is based on the behavioral change observed in infected individuals. They typically spend a significant amount of time scratching themselves, diverting their attention from other activities, reducing its performance and fitness  \citep{arzamendia2012effect, escobar2022sarcoptic,martin2018cascading, sosa2022occurrence}. Based on this, in our model, infected individuals do not contribute to population growth, while still consume their share of resources, as reflected in Eq.~(\ref{eq1S}). 

We have analyzed the steady state solutions of the system of equations~(\ref{eq1S}-\ref{eq1F}), as well as their stability. We have also performed an extensive numerical study of the dynamical system, including a bifurcation analysis for each steady state solution of the system as a function of the fomite control parameters $\beta_{f}$ and $\rho$.

\section{Results}

\subsection{Steady state solution analysis}
The SEI system, Eqs.~(\ref{eq0S}-\ref{eq0I}), has 2 relevant equilibria (besides the extinction one, $S=E=I=0$, which is always an unstable state, and one with negative populations). The phase space is organized in two regimes: a disease-free state and an endemic one. An appropriate control parameter to characterize the transition between them is the ratio between the infection rate and the infection-related death rate, $R_0 = \beta_1 /\mu$, which can be interpreted as the basic reproductive number of the epidemic, since it represents the number of contagions produced by each infected, while alive. As expected, if the death rate is faster than the contagion (that is, if $R_0 <1$) the system is free of disease. On the contrary, if $R_0 >1$, that state is unstable and the system displays a persistent infected population. The disease-free state is a stable node and only attractor of the dynamics, when $R_0<1$. At $R_0=1$ there is a transcritical bifurcation, where the two equilibria interchange stability, and the the system enters an endemic state when $R_0>1$. 

\begin{figure}[t]
\centering
\includegraphics[width=\columnwidth]{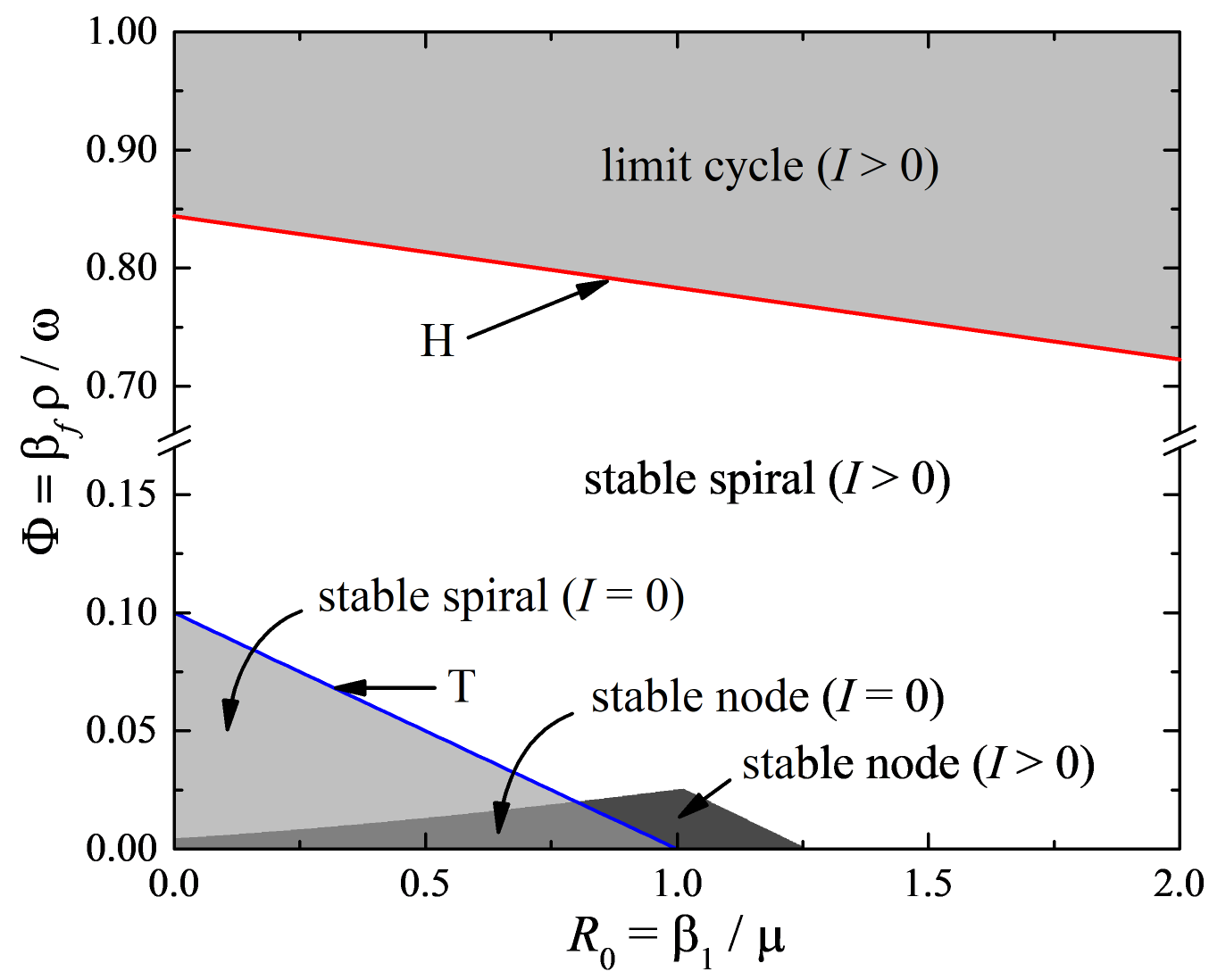}
\caption{Phase diagram of the SEIF system. The bifurcation lines where identified through the eigenvalues of the linearized system. The bifurcation marked T is transcritical (blue line), while the other one is Hopf (red line). Note that there is a break in the vertical axis. Parameters: $\mu=0.1$, $\beta_f=0.5$, $\omega=0.4$, $r=0.11$, $\gamma=0.5$, while  $\beta_1$ and $\rho$ where varied to cover the range of control parameters.
}
\label{fig:phases}
\end{figure}

The SEIF model also features two significant equilibria which are analogous to those in the SEI model. However, the stability of these equilibria cannot be assessed by considering only the previously defined $R_0$, as it is the basic reproductive number of a system that considers only direct transmission. One way to describe the behavior of the SEIF model is by considering an additional control parameter, $\Phi$ = $\beta_f \rho / \omega$, which characterizes the dynamics of fomites. This allows us to draw a phase diagram (see Fig.~\ref{fig:phases}), showing the stable states of the system based on two control parameters, $R_0$ and $\Phi$. In this representation, the SEI system is restricted to the horizontal axis, where either $\beta_f$ or $\rho$ are zero.\footnote{See mathematical details in the Supplementary Material.}

To isolate their effects on the steady state of the system, we kept the remaining parameters fixed. The growth rate will be kept as $r = 0.11$, corresponding to the system of wild camelids that inspired our model \citep{puig2000dinamica}, whereas  $\gamma = 0.5$ corresponds to a incubation time of a couple of days. Their exact values do not affect the structure of the phase space of the model that we will describe below.

The phase diagram shows that the disease-free state ($I=0$) is restricted to a triangular region around the origin. Increasing the role of the fomites destabilizes this state, which bifurcates into an endemic state with $I>0$. The transcritical bifurcation (marked T in Fig.~\ref{fig:phases}) can be found analytically, and it is given by the straight line:
\begin{align}
    \Phi = \mu \, (1-R_0).
\label{eq:transcritic}    
\end{align}
 Figure~\ref{fig:phases} shows that this transition is rather involved. The triangular disease-free region comprises a stable node (adjacent to the horizontal axis, which is the SEI system) and a stable spiral that develops from it when a pair of eigenvalues become complex.\footnote{Bear in mind that the space is four-dimensional, which allows a spiral dynamics yet with $I=0$.} The transcritical line T, shown in blue in Fig.~\ref{fig:phases}, separates both these states from corresponding ones with and endemic ($I>0$) state. Most of this phase consists of a stable spiral, but a small rhomboidal region is actually a stable node, which is also adjacent to the corresponding state in the SEI system (the horizontal axis).

The structure of the phase diagram of Fig.~\ref{fig:phases} highlights the crucial role of the fomites as a reservoir of disease: they allow the infectious agent to avoid extinction. We see that, even for $R_0<1$ (which, in the SEI system, ensures the disappearance of the disease), if the SEIF system lies above the transcritical line T, the infection persists.
The key role of the fomites manifests itself also at larger values of the control parameters. There is a further transition that destabilizes the spiral and produces a limit cycle as a transcritical Hopf bifurcation (H). These transitions will be further elaborated on in the following. 

\begin{figure}[t]
\centering
\includegraphics[width=\columnwidth]{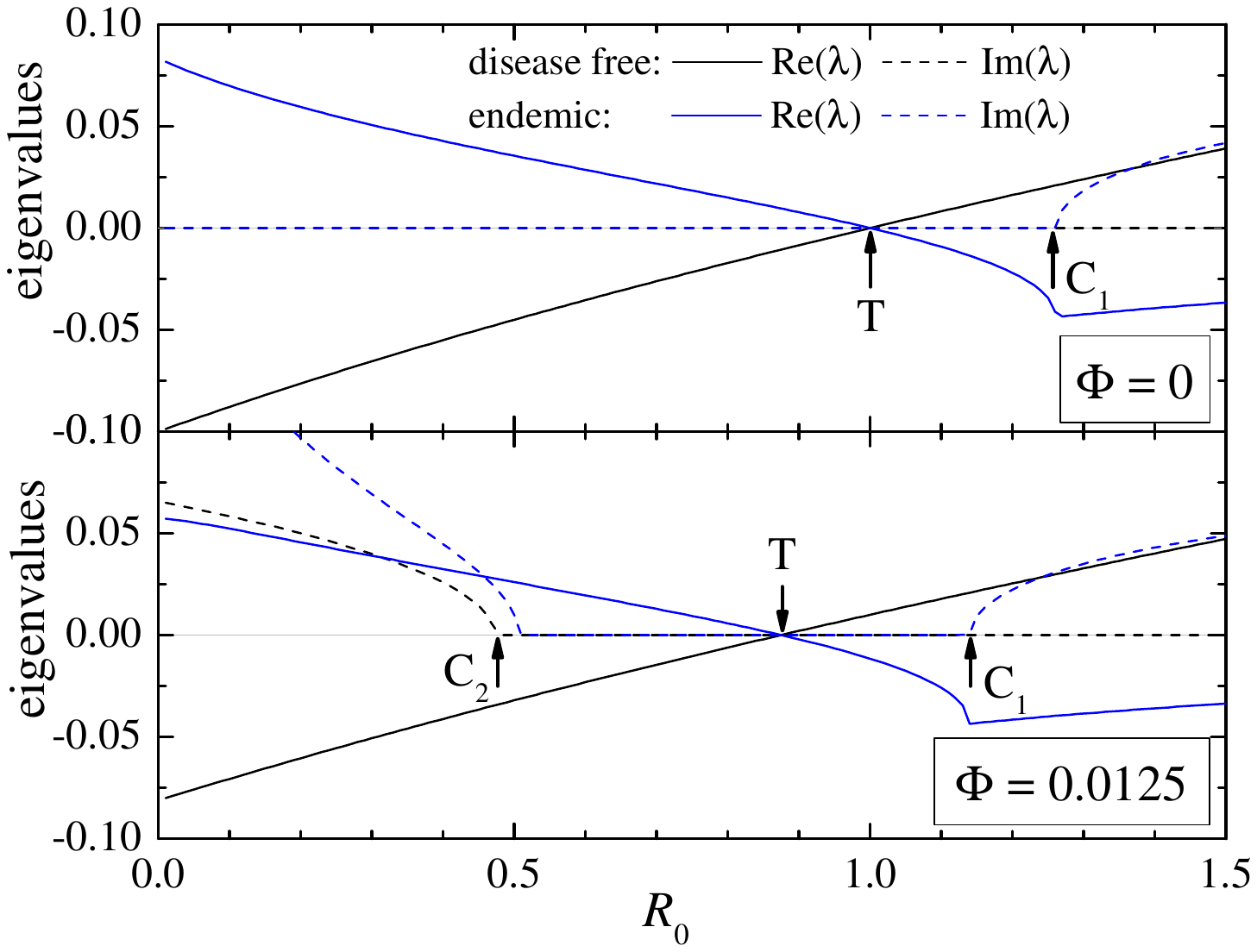}
\caption{Real and imaginary parts of the eigenvalues controlling the stability of the disease-free ($I=0$, top panel) and the endemic ($I>0$, bottom panel) phases. Each diagram corresponds to a different cut on the vertical axis ($\Phi$) of the phase diagram of Fig.~\ref{fig:phases}. At the point T, the real part of an eigenvalue of the disease-free state exchange sign with the one of the endemic state, which indicates a change of stability on both equilibrium points. Additionally, C$_1$ and C$_2$ indicate the points where their respective eigenvalues become real or complex, respectively. Parameters: $r=0.11$, $\gamma=0.5$, $\mu=0.1$, and for the SEIF system: $\beta_f=0.5$, $\omega=0.4$, $\rho=0.01$ ($\beta_1$ was varied to cover the range of the control parameter $R_0$).}
\label{fig:eigenvalues1}
\end{figure}

A straightforward approach to describing the system's behavior is through horizontal and vertical cross-sections of the phase diagram just presented.
In Fig.~\ref{fig:eigenvalues1}, we illustrate two horizontal sections corresponding to the cases $\Phi = 0$ (SEI model, without fomites) and $\Phi =0.0125$ (SEIF model), depicting the behavior of the relevant eigenvalues as functions of $R_0$.

The relevant eigenvalue of the disease-free state (Fig.~\ref{fig:eigenvalues1}, top panel) is shown in black (full line is the real part, while the imaginary part is dashed), and the one corresponding to the endemic state is blue (same convention for real and imaginary parts). The critical point, marked T, is the transcritical bifurcation between both states, and the real parts cross at the value 0, interchanging stability. It can also be seen that, while the disease-free phase is always a stable node (null imaginary part), the endemic state becomes a stable spiral at a larger value of $R_0$. This is analogous to a mechanical oscillator at the critical damping, changing from overdamped to underdamped,  and consequently we marked this point as C$_1$ in the graph. 

The addition of fomites, as represented by Eqs.~(\ref{eq1S}-\ref{eq1F}), preserves and extends this structure in the larger phase space. The stability of the two relevant equilibria can be assessed analytically. The real and imaginary parts of the relevant eigenvalues, for a representative case, are shown in Fig.~\ref{fig:eigenvalues1} (bottom panel). The transcritical bifurcation (T) is still there, as well as the transformation of the endemic node into a spiral (C$_ 1$), but there is an additional critical point in the region where the disease-free state is stable, from which the spiral becomes a node (C$_2$). 

\begin{figure}[t]
\centering
\includegraphics[width=\columnwidth]{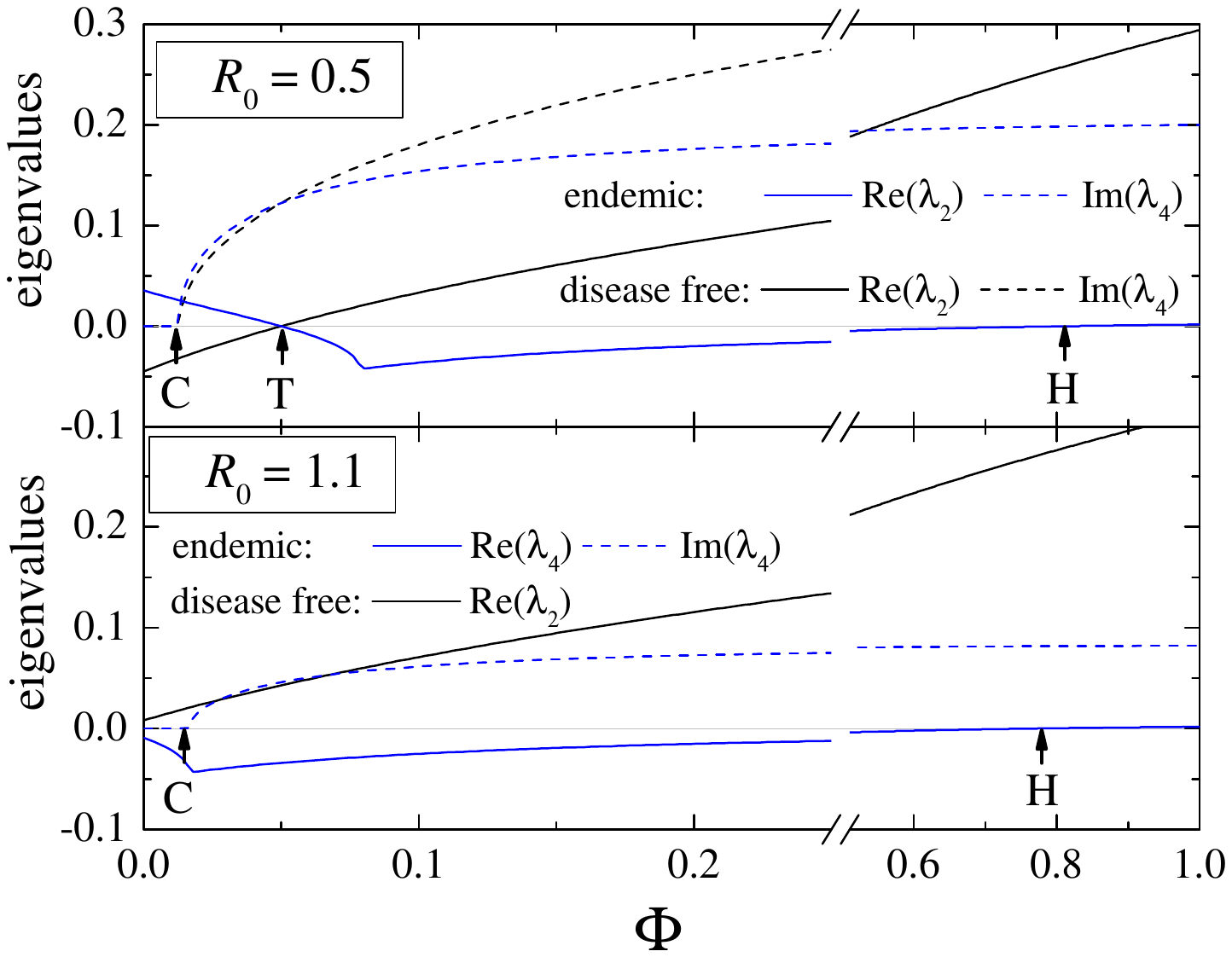}
\caption{Real and imaginary parts of the eigenvalues controlling the stability of the disease-free ($I = 0$, top panel) and the endemic ($I > 0$, bottom panel) phases. Note that there is a break in the horizontal axis. Each diagram corresponds to a different cut on the horizontal axis ($R_0$) of the phase diagram, as indicated. The point H marks a Hopf bifurcation and the onset of a limit cycle. The point C indicates the point where the eigenvalues become complex. Parameters: $\beta_1 = 0.05$ and 0.15, $\mu=0.1$, $\beta_f=0.5$, $\omega=0.4$, $r=0.11$, $\gamma=0.5$, while  $\rho$ was varied to cover the range of the control parameter.}
\label{fig:eigenvalues2}
\end{figure}

To complete the analysis, we display in Figure~\ref{fig:eigenvalues2} the relevant eigenvalues corresponding to vertical cuts at values $R_0=0.5$ and $R_0=1.1$ of the diagram presented in Fig.~\ref{fig:phases}, as functions of $\Phi$. 
At large values of $\Phi$ a transcritical Hopf bifurcation (H) takes place, as the real part of one of the eigenvalues of the endemic state turns positive, resulting in the loss of stability of that equilibrium point. At small values of $\Phi$, on the other hand, a transition C is observed where the eigenvalues become complex.

\begin{figure}[t]
\centering
\includegraphics[width=\columnwidth]{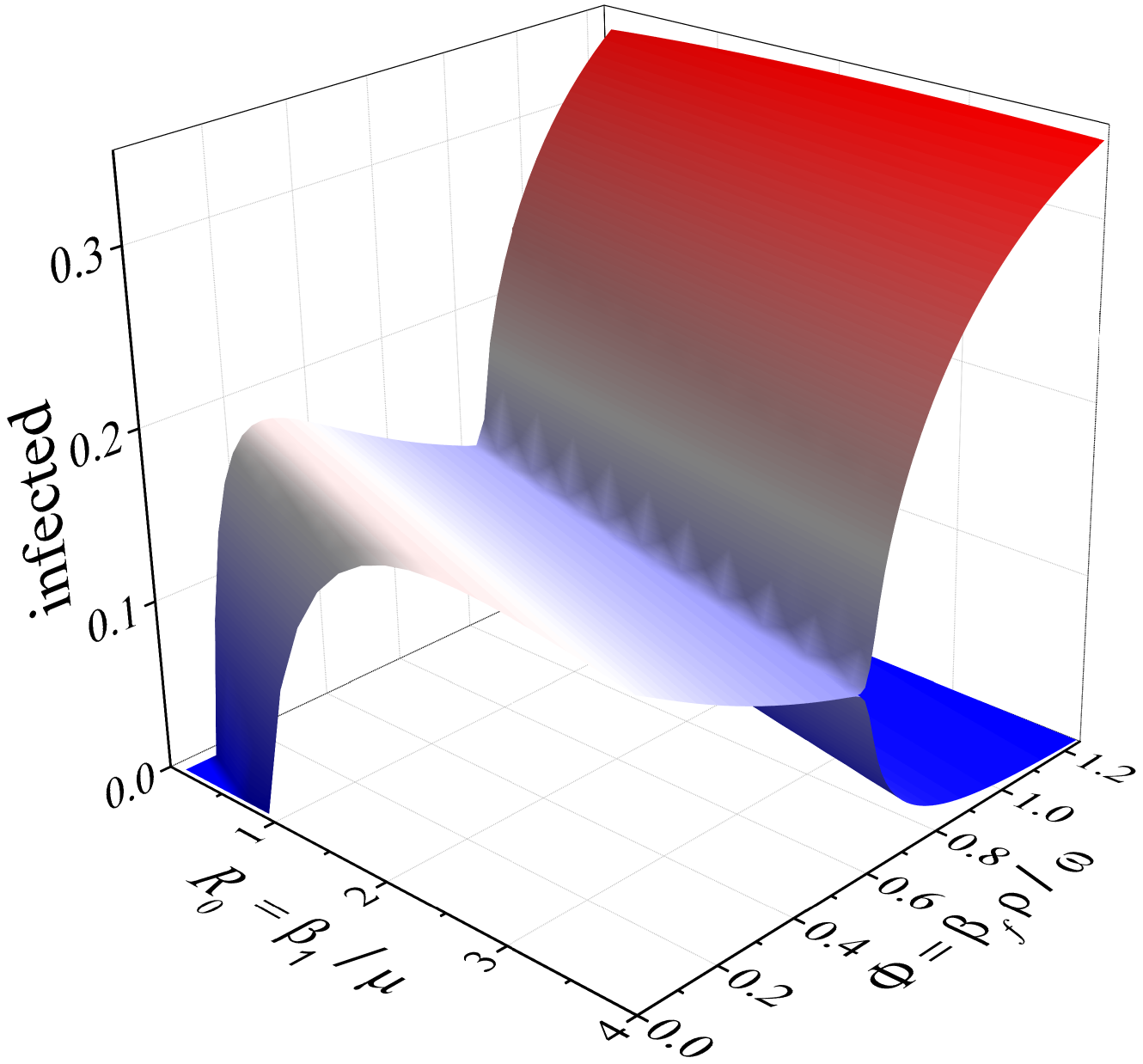}
\caption{
Bifurcation diagram, showing the infected population size projected from the $R_0$-$\phi$ plane.}
\label{fig:bifdiag}
\end{figure}

To complement the visualization of the phases shown in Fig.~\ref{fig:phases}, we present a bifurcation diagram in Fig.~\ref{fig:bifdiag}, using the infected population as order parameter. The phase of no infection can be seen, again, as a triangle near the origin. The stable center of the node or spiral occupies the adjacent phase. In the region of the limit cycle, we show the maximum and minimum of the infection oscillation as order parameters. 

There are several interesting features of the scenario presented by these results. Firstly, it is worth noting the way in which the amplitude of the cycles grows, for increasing values of the fomite parameter (which means increasing $\beta_f$ or $\rho$, or decreasing $\omega$).

Another interesting feature is the non-monotonic behavior observed in the bifurcation diagram. There exists a broad range across both control parameters where the infected phase decreases as the parameters governing the infection severity increase (Fig. \ref{fig:bifdiag}). Nevertheless, since the model includes a vital dynamics that also affects the susceptible population, the fraction of infected always grows (not shown here). 

\bigskip

\section{Discussion}

\subsection{An endemic state induced by indirect transmission}

\begin{figure}[t]
\centering
\includegraphics[width=\linewidth]{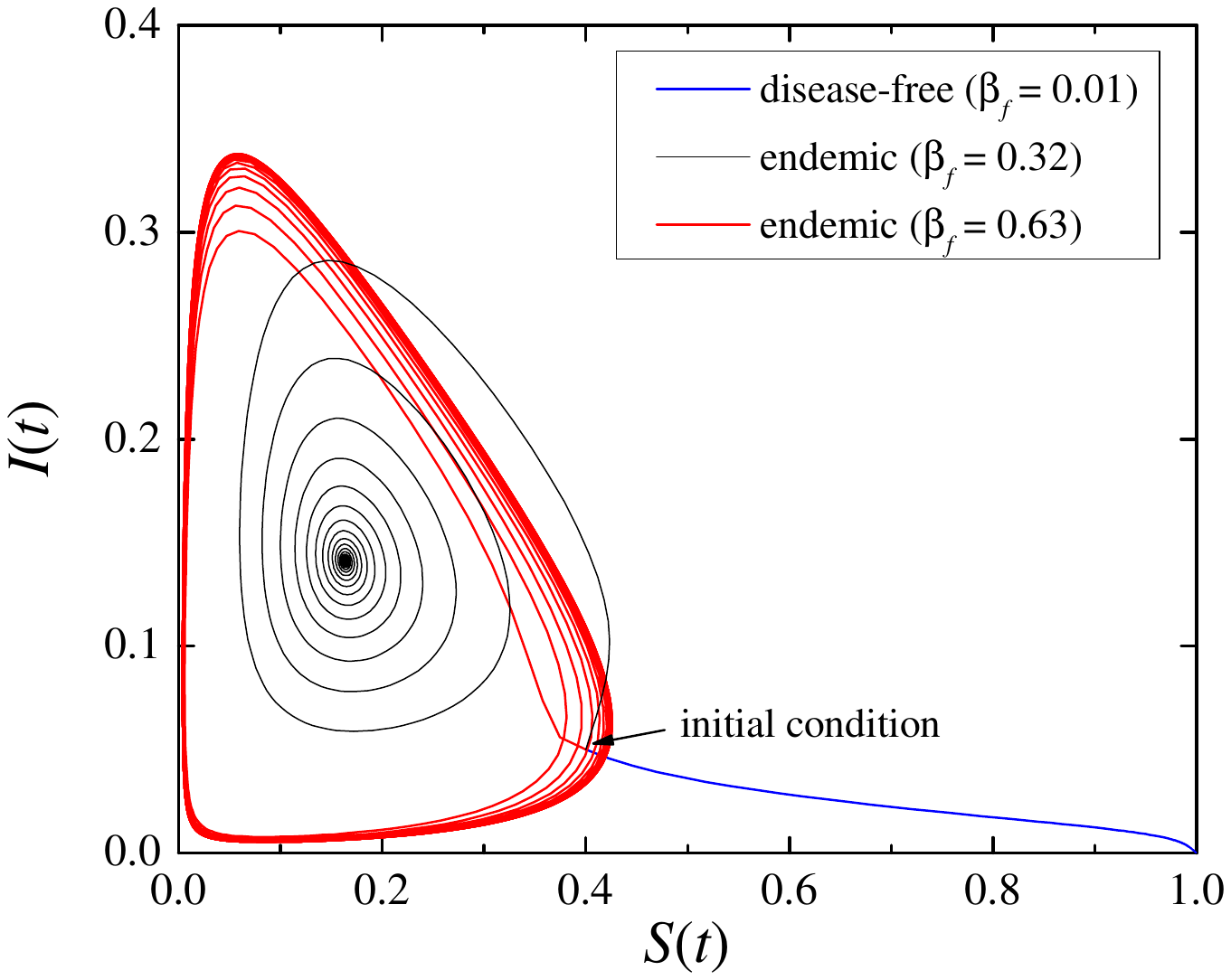}
\caption{Trajectories in the reduced susceptible-infected phase space. When $\Phi$ is subcritical, the system stabilizes in a disease-free stable node (blue line). When $\Phi$ is  below H, the stable state is an endemic spiral. When it is beyond the Hopf bifurcation, the stable state is an endemic limit cycle (red line). The trajectories were obtained for three different values of $\beta_f$ as indicated in the text, while keeping the rest of the parameters fixed at the values $r = 0.11$, $\beta_1 = 0.05$, $\gamma = 0.5$, $\mu = 0.1$, $\rho = 0.7$, and $\omega = 0.4$.}
\label{fig:trajectories}
\end{figure}

As described in the previous section and shown in Fig.~\ref{fig:bifdiag}, a fomites-free system (the horizontal axis of Fig.~\ref{fig:phases}, where either $\beta_{f}$ or $\rho$ are zero), with $R_0$ smaller than 1, stabilizes at the disease-free state. If there are fomites, but they are ephemeral as defined by $\Phi$, the same situation holds. However, with the same value of $R_0$ (e.g., same contagion and death rates), the system can abandon the disease-free state, and instead stabilize at the endemic state on a spiral, if the following condition is fulfilled:  
\begin{equation}
\ \Phi > \mu - \beta_1.
\label{eq6}
\end{equation}
This threshold corresponds to the point T in the Fig. ~\ref{fig:eigenvalues1}, where there is change of sign in the eigenvalues of the endemic state that are sensitive to fomite parameters. With a further increase of $\Phi$, the spiral loses its stability and a stable limit cycle is formed around the equilibrium. This assessment of the speed of environmental transmission was also made at \citep{cortez2013distinguishing}, but here we were able to determine the specific minimum level for successful disease invasion.
The figure labeled Fig. \ref{fig:trajectories} illustrates this concept, with the blue line representing the system trajectory when $\Phi$ is below this minimum level, stabilizing in the disease-free state, while trajectories depicted by the black and red diverge when $\Phi$ exceeds the minimum level, showcasing stable spiral and limit cycle dynamics discussed in previous section.

The implications of this result are most interesting, since it shows that the disease persists under a regime where the infected individuals die faster than they would be replaced on the classical SEI system with no indirect transmission. The fomites can be thought of as an environmental cache of the infectious agent, that enhance the chances of having an endemic state  by increasing the ``contagion risk'' of the susceptible population, almost independently of the infected population itself. Although those infected tend to be eliminated rapidly, they can spread the infection postmortem through these reservoirs.

Once the system enters the phase of sustained oscillations, as shown in Fig.~\ref{fig:bifdiag}, the infection recovers from a small value at each successive cycle. Such behavior is reasonable in a continuous model. However, it should be noted that an actual system would consist of discrete individuals and be subject to stochastic factors (either demographic or environmental). In such a case, it is unavoidable that the infection would disappear due to fluctuations. Without an external source of infected, the system would remain free of infection.

Based on bibliographic data, we decided to exclude the infected individuals from the reproductive process. For a further test of the implications of this assumption, we performed a variant of our model Eq.~(\ref{eq1S}-\ref{eq1F}) including the infected fraction in the vital term. This resulted on the vanishing of the limit cycle phase and more damped oscillations in the spiral phases compared to the previous case (not shown here).
This outcome is interesting, considering that in the similar system analyzed in Ref.~\citep{cortez2013distinguishing}, no analytical differences were found in the system under different growth functions.

\begin{figure}[t]
\centering

\includegraphics[width=\linewidth]{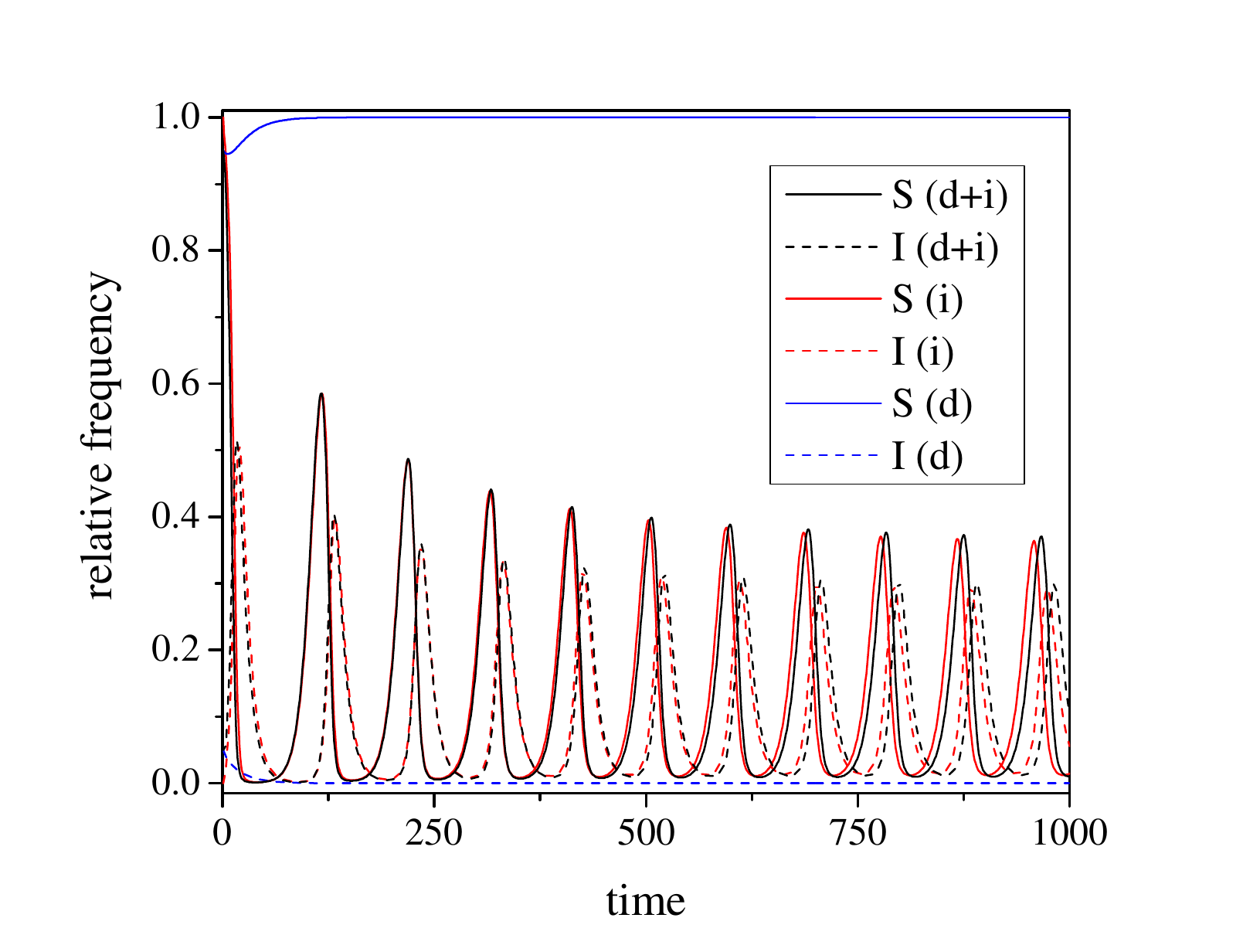}
\caption{Time series of disease evolution under different transmission mechanisms under the regime $R_0$ $<$ 1. Starting from the same initial condition, the system stabilizes at the disease-free equilibrium in the absence indirect transmission to reinforce the disease spread (blue lines). In cases involving indirect transmission, whether with direct contact (black lines) or even without it (red lines), the disease can persist within the system.} 
\label{fig:Fig6}
\end{figure}

\subsection{Outbreak trigger}

In the same regime ($R_0<1$), an outbreak can begin after a small perturbation from the disease-free state, unlike the situation of a system without fomites. 
It can even arise and persist without any level of direct transmission ($\beta_{1} = 0$), as illustrated in Fig. \ref{fig:Fig6}, in the series labeled as $I(i)$. It's also noteworthy that, even without direct transmission, the disease can enter the endemic state with significant oscillations in $S(i)$ and $I(i)$, that regularly bring the population close to extinction.

Additionally, we note the striking resemblance between the series labeled as $I(d)$ (only direct transmission) and $I(d+i)$ (both direct and indirect transmission).

Using the threshold expressed by Eq.~(\ref{eq6}), which corresponds directly to the  transcritical line T marked in Fig.~\ref{fig:phases}, we can rewrite it to define an effective reproductive number for the system with indirect transmission, $R_S$,  as a replacement of the now inadequate $R_0$. This parameter would be:
\begin{align}
R_S = \frac{1}{\mu} \left(\beta_{1} + \frac{\rho\,\beta_{f}}{\omega}\right).
\end{align}

This parameter has been calculated in Ref.~\citep{yagci2019comparing}; however, an important distinction is that our system does not include a term for the replacement of the susceptible population. Considering this main difference, it is remarkable that we still can predict if the system goes to the disease-free or the endemic states, based on such a simple generalization of $R_0$. Even when there is no substitution or recovery of the infected population, if $R_S \leq$ 1, the disease is extinguished, otherwise it persists.

As we can observe in Fig.~\ref{fig:Fig6}, there is a small difference between the system with only indirect transmission (curves of $S(i)$ and $I(i)$) and a system with both forms of transmission (curves of $S(d+i)$ and $I(d+i)$). This is related to the fact that the $R_S$ values between these curves are very close: $R_S = 11$ for direct transmission and $R_S = 10.5$ for the mixed case.

\subsection{Conclusion}

Our results show that the indirect transmission by fomites can actually perform as an outbreak trigger and as an effective stable reservoir of the infectious agent, allowing the disease to enter into an endemic state, even under a regime where the infected individuals die faster than they transmit the disease to others. 
These outcomes are similar to those observed in other systems \citep{yagci2019comparing, cortez2013distinguishing}, but conducting a more detailed study of the equilibrium points allowed us to obtain a more precise description of the phase space and further classification of the different regions shown in Fig.~ (\ref{fig:phases}).

It has been described that, in many human and wildlife populations, sarcoptic mange persists as an endemic disease, generating periodic fluctuations of the population. It has also been observed that, in the initial waves, the populations experiment high rates of mortality, and that in later waves the mortality rate is reduced, arguing  the selection of resistant individuals and herd immunity \citep{perez2021biology}. As we showed in this work, this damped oscillations could also emerge from the mathematical properties of the system, without involving selection or immune response processes (we did not even considered a recovery compartment in our model). 

Mathematical modeling allows us to describe the relation between the different variables and different courses of the disease, which is a useful tool to evaluate possible measures of intervention or management of the disease \citep{garnett2011mathematical, keeling2011modeling, anderson1991infectious}. For wildlife, the most considered measures are eradication, control and prevention \citep{perez2021biology,wobeser2013investigation}. Based on our model and results, the eradication of infected individuals does not necessarily put an end to the outbreak, given the fact that, in the SEIF system, there is a second source of infection in the form of fomites, which makes the system very vulnerable to small perturbations. The control of the spread of a disease by preventing the direct transmission between hosts, which in our model is represented by the parameter $\beta_1$, could also not be enough, considering that there is another type of contagion quantified by $\beta_f$, and the disease persists even without any direct transmission ($\beta_{1}$ = 0, Fig.~\ref{fig:Fig6}). For these reasons we consider that the best kind of measure would be prevention, which implies reducing the probability of encounter of a disease reservoir and to get the pathogen (which in terms of our model would mean reducing $\beta_{f}$), and therefore moving the system away for the endemic phase. 

Nevertheless, even if our fomite parameters ($\rho$, $\beta_{f}$ and $\omega$) capture many factors involving the parasite's fitness on the system off-host, it is important to remember that the mean field approximation does not take into consideration other relevant phenomena, such as the host species social and foraging behavior or the structure of the habitat \citep{escobar2022sarcoptic,browne2022sustaining}. Some of them would bring a stochastic aspect into the mathematical modelling, as mentioned above, or modifications to the well-mixed assumption inherent to the mass-action equations analyzed in the present work. All these factors could lead to outcomes different from those predicted here. For instance, stochastic or seasonal behavior or the inclusion of the Allee effect \citep{kuperman2021allee}, along with the oscillating nature of the endemic state, could result in the total extinction of the population, a state that, as mentioned, is unstable in our mean-field model. These aspects are currently been taking into consideration in the development of a spatially explicit model, where we will test the relative importance of every fomite parameter in the evolution of a sarcoptic mange infection on a more realistic scenario, inspired by the outbreak that took place in  the wild camelids populations of San Guillermo National Park \citep{ferreyra2022sarcoptic,monk2022}. 

Still, it is worth stressing the biological significance of  fomites and their potential role as an outbreak trigger and disease reservoir, considering the fact that several of the parameters considered (such as $\rho$ and $\omega$) are functionally related only to traits that belong to the parasite, and not to traits of the host. With this in mind, if we consider that the real system could have any similarity to our mean field model (see, for example, \citep{beeton2019model}), the parasite could improve its fitness by enhancing characteristics that are beyond co-evolution processes with the host. This kind of betterment is natural of parasites in wildlife, as they are under strong selective pressures to evolve strategies that enable them to persist \citep{browne2022sustaining,hudson2002ecology,swinton2002microparasite}.

\section{Acknowledgements}
This research was supported by Agencia Nacional de Promoción Científica y Tecnológica (PICT 2018-01181, PICT 2019-02167 and PICT
2020–0875), Consejo Nacional de Investigaciones Científicas y Técnicas (CONICET, PIP 112-2022-0100160 CO)  and Universidad Nacional de Cuyo (06/C045-T1). 

The authors thank M. N. Kuperman for invaluable discussions. 

G. Abramson would like to thank the Isaac Newton Institute for Mathematical Sciences, Cambridge, for support and hospitality during the programme Mathematics of movement: an interdisciplinary approach to mutual challenges in animal ecology and cell biology, where work on this paper was undertaken. This work was supported by EPSRC grant no EP/R014604/1.

\bibliographystyle{ieeetr}
\bibliography{ref} 

\end{document}

\begin{frontmatter} 
\title{% 
A mean field analysis of the role of indirect transmission in emergent infection events \\
\vspace{0.5cm}
\underline{Supplementary material}}

\author[tomas]{Tomás Ignacio González\corref{cor1}}
\ead{tomignaciogon@gmail.com}
\cortext[cor1]{Corresponding author}

\author[fabiana]{María Fabiana Laguna}
\ead{lagunaf@cab.cnea.gov.ar}

\author[guillermo]{Guillermo Abramson}
\ead{guillermo.abramson@ib.edu.ar}

\address[tomas]{Statistical and Interdisciplinary Physics Division, Centro Atómico Bariloche (CNEA), R8402AGP Bariloche, Argentina.}

\address[fabiana]{Statistical and Interdisciplinary Physics Division, Centro Atómico Bariloche (CNEA) and CONICET. Universidad Nacional de Río Negro. R8402AGP Bariloche, Argentina.}

\address[guillermo]{Statistical and Interdisciplinary Physics Division, Centro Atómico Bariloche (CNEA), CONICET and Instituto Balseiro (Universidad Nacional de Cuyo). R8402AGP Bariloche, Argentina.}

\end{frontmatter}

This document provides additional details of the stability analysis used in our study. While the exact expressions and values for the fixed points and parameters are not shown, our intention is to indicate the technique used during the development of our work.   

\section{Calculation of the fixed points}
We begin by calculating the fixed points of the phase space, where the system reaches equilibrium. 

Initially, we performed an analysis over our reference SEI system. 

\begin{align}
\frac{dS}{dt} &= F(S,E,I) = r(S+E)(1-S-E-I) - \beta_1 SI,\\
\frac{dE}{dt} &= G(S,E,I) = \beta_1 SI  - \gamma E,\\
\frac{dI}{dt} &= H(S,E,I) = \gamma E - \mu I.
\end{align}

In this case, it entails seeking the equilibrium values $S^*$, $E^*$, $I^*$ defined as the solutions to the following condition:
\begin{align}
    F(S^*,E^*,I^*) = G(S^*,E^*,I^*) = H(S^*,E^*,I^*) = 0.
\end{align}
By solving this equations, we obtain several sets of equilibrium values for each variable. In the case of the SEI system, there are four distinct fixed points. These are the extinction equilibrium ($S^*= 0$, $E^*= 0$, $I^*= 0$), the disease-free equilibrium ($S^*= 1$, $E^*= 0$, $I^* = 0$), and two non-trivial equilibria whose expressions are too involved to determine analytically and require mathematical software. For our study, we performed the corresponded calculus using Wolfram Mathematica. 

After conducting a numerical analysis of these latter two equilibria, we observe that one of them results in non-positive values, while the other defines an endemic equilibrium for the disease ($S^* >0$; $E^* > 0$, $I^* >0$). We consider the non-positive equilibrium as meaningless and only take into consideration the other three for further analysis.

\section{Jacobian matrix}

Having obtained the three fixed points of interest, we proceed to determine the stability of each one, using the Jacobian matrix, which is defined as:

\begin{align*}
J = 
\begin{pmatrix}
\frac{dF(S,E,I)}{dS} & \frac{dF(S,E,I)}{dE} &\frac{dF(S,E,I)}{dI} \\
\frac{dG(S,E,I)}{dS} & \frac{dG(S,E,I)}{dE} &\frac{dG(S,E,I)}{dI} \\
\frac{dH(S,E,I)}{dS} & \frac{dH(S,E,I)}{dE} &\frac{dH(S,E,I)}{dI} 
\end{pmatrix}.
\end{align*}

By evaluating the Jacobian matrix with the respective equilibrium values of each variable $(S^*,E^*,I^*)$, and then calculating its eigenvalues, we can determine not only the stability of the equilibrium but also the behavior of the trajectories in the neighboring region. 

\section{Example}
Lets consider the extinction equilibrium. First, we calculate the Jacobian matrix:
\begin{align*}
J = 
\begin{pmatrix}
-\beta_1 I - r (-1 + 2E + I + 2S) & -r (-1 + 2E + I + 2S) & -E r - (\beta_1 + r) S \\
\beta_1 I & -\gamma &\beta_1 S \\
0 & \gamma &-\mu
\end{pmatrix}
\end{align*}
Second, we evaluate the matrix at the corresponded values (S* = 0; E* = 0; I* = 0):
\begin{align*}
J(0,0,0) = 
\begin{pmatrix}
r & r & 0 \\
0 & -\gamma & 0  \\
0 & \gamma &-\mu 
\end{pmatrix}.
\end{align*}
The corresponding set of eigenvalues for this matrix are:
\begin{align*}
    \lambda = (r,-\mu, -\gamma).
\end{align*}

%\section{Eigenvalue interpretation}
Because every eigenvalue of this set belongs to the domain of real numbers, we can determine that this fixed point is a node. Additionally, since one of the eigenvalues ($r$) is always a positive number, we conclude that it is an unstable node. This implies that after a slight perturbation from the fixed point, the perturbation increases, and the trajectories diverge from the extinction equilibrium.

\section{Further delevopment: the SEIF model}

We performed this same procedure for all other equilibria of interest from the SEI model and also for the ones corresponding to the SEIF model (with the difference being that this model is four-dimensional instead of three-dimensional).

Our SEIF system has a similar set of fixed points to those of the SEI classic model: an extinction equilibrium, a disease-free one, an endemic disease state, and a non-interesting non-positive equilibrium. 

In the Results section of the main manuscript, we focus on the disease-free and the endemic equilibria, which showed to be actually affected by fomite parameters and, as such, of our most interest. Using this method of analysis, we could describe a new set of possible dynamics of the disease that were beyond the predictive capacity of the classical $R_0$ parameter. Read the full manuscript for further results and discussion.

\bibliographystyle{ieeetr}